\documentstyle[amssymb,graphicx]{article}
\begin{document}
\title{Formation time of hadrons in photon induced reactions on nuclei\footnote{Work supported by DFG.}}

\author{T. Falter and U. Mosel\\
\small{\it Institut fuer Theoretische Physik}\\ 
\small{\it Universitaet Giessen}\\ 
\small{\it D-35392 Giessen, Germany}}

\date{February 11, 2002}

\maketitle

\begin{abstract}
We present a way to account for coherence length effects in a 
semi-classical transport model. This allows us to describe photo- and 
electroproduction at large nuclei ($A>12$) and high energies using a realistic
coupled channel description of the final state interactions that goes far 
beyond simple Glauber theory. We show that the purely absorptive treatment of 
the final state interactions as usually done in simple Glauber theory might 
lead to wrong estimates of color transparency and formation time effects in 
particle production. As an example we discuss exclusive $\rho^0$ 
photoproduction on Pb at a photon energy of 7~GeV as well as $K^+$ and $K^-$ 
production on C and Pb in the photon energy range 1-7~GeV.
\end{abstract}
\bigskip

\section{Introduction} \label{sec:intro}
In a high energy collision between two hadrons or a photon and a hadron it 
takes a finite amount of time for the reaction products to evolve to physical 
particles. During the collision process some momentum transfer between the 
hadrons or some hard scattering between two of the hadrons' constituents 
leads to the excitation of hadronic strings. The time that is needed for the 
creation and fission of these strings as well as for the hadronization of the
string fragments cannot be calculated within perturbative QCD because the 
hadronization process
involves small momentum transfers of typically only a few hundred MeV. One can 
perform an estimate of the formation time $\tau_f$ in the rest frame of the 
hadron. It should be of the order of the time that the quark-antiquark 
(quark-diquark) pair needs to reach a separation that is of the 
size of the produced hadron ($r_h\approx 0.6-0.8$~fm):
\begin{equation}  
  \tau_f\gtrsim\frac{r_h}{c}.\\
\end{equation}

During their evolution to physical hadrons the reaction products will react 
with reduced cross sections. This is motivated by means of color transparency: 
the strings and the substrings created during the fragmentation are in a color 
singlet state
and therefore mainly react via their color dipole moment which is proportional 
to their transverse size. For a collision inside nuclear medium this means that
during their formation time the produced hadrons travel with a reduced 
scattering probability. Hence the formation time plays an important role in the
dynamics of nuclear reactions, e.g. heavy ion collisions, proton and pion 
induced reactions as well as photon and electron induced reactions on nuclei.
The latter two are of special interest because they are less complex than 
heavy ion collisions and instead of hadron induced reactions the primary 
reaction does in general not only take place at the surface of the nucleus but 
also at larger densities. Experiments at TJNAF and DESY, for example, deal with 
exclusive and inclusive meson photo- and electroproduction at high energies. 
Large photon energies $E_\gamma$ are of special interest because the formation 
length $l_f$ in the rest frame of the nucleus can exceed nuclear dimensions:
\begin{equation}
  l_f=v_h\cdot\gamma\cdot\tau_f=\frac{p_h}{m_h}\cdot\tau_f.
\end{equation}
If one chooses the formation time to be $\tau_f=0.8$~fm/c, the formation length
\begin{figure}[bt]
  \begin{center}
    \includegraphics[width=8cm]{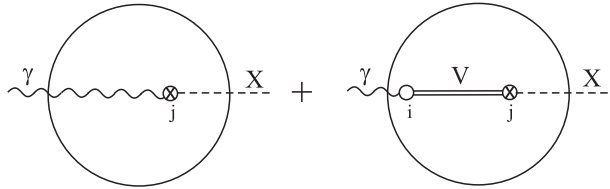}
  \end{center}
  \vspace{-0.2cm}
  \caption{The two amplitudes of order $\alpha_{em}$ that contribute to 
incoherent meson photoproduction in simple Glauber theory. The left amplitude
alone would lead to an unshadowed cross section. Its interference with the
right amplitude gives rise to shadowing.}
  \label{fig:Glauber}
\end{figure}  
in the rest frame of the nucleus will be about 30~fm for a 5~GeV pion and about
7~fm for a 5~GeV Kaon or a 7~GeV $\rho$ meson. These lengths have to be
compared with the typical size of nuclear radii, e.g. 2.7~fm for $^{12}$C and 
7.1~fm for $^{208}$Pb. The formation time has therefore a big effect on 
photonuclear production cross section at high energies.\\ 

To extract informations 
about the formation time one needs a realistic description of the final state 
interactions (FSI) of the reaction products. Since photon induced reactions 
are known to be shadowed ($\sigma_{\gamma A}<A\sigma_{\gamma N}$) above 
$E_\gamma\approx 1$~GeV~\cite{Bia96Muc99,Fal00,Fal01}, one also needs a way to
account for this shadowing effect in photoproduction. This is straight forward
within Glauber theory~\cite{Gla59Yen71} 
but in simple Glauber theory the FSI are purely absorptive. A more 
realistic coupled channel description of the FSI is possible within a 
transport model. We use a semi-classical transport model based on the 
Boltzmann-Uehling-Uhlenbeck (BUU) equation to describe the FSI.
Originally developed to describe heavy ion 
collisions~\cite{SIS} at SIS energies it has been extended 
in later works to investigate also inclusive particle production in heavy ion 
collisions up to 200~AGeV and $\pi$~\cite{Eff99a} and $p$ induced as well as 
photon and electron induced reactions in the resonance 
region~\cite{res-reg}. Inclusive photoproduction of mesons at energies between 
1 and 7~GeV has been investigated in~\cite{Eff00}. An attractive feature of 
this model is its capability to describe a large variety of different reaction
types.
However it is not clear how to account for coherence length effects such as 
shadowing in a semi-classical transport model. A first attempt has been made 
in~\cite{Eff00}, in the next section we present a new, better way to implement 
shadowing in our model.

\section{Model}
Within our model we can only 
calculate the incoherent part of the photon nucleus cross section and we use 
Glauber theory to calculate coherent processes such as, e.g., coherent vector 
meson 
photoproduction. Note that the main part of the difference between the nuclear 
photoabsorption cross section and the incoherent photoproduction cross section 
at large energies stems from coherent $\rho^0$ photoproduction.
For a detailed discussion of the used transport model and the way how to
implement the shadowing effect we refer to \cite{Eff99b} and \cite{Fal02}.
In our model the reaction of a high energy photon with a nucleus takes place in
two steps. In the first step the photon reacts with one nucleon inside the 
nucleus (impulse approximation) and produces some final state $X$. In this 
process nuclear effects like Fermi motion, binding energies and Pauli blocking
of the final state nucleons are taken into account. In the second step the 
final state $X$ is propagated within the transport model. 
Except for the exclusive vector meson and exclusive strangeness production 
(see~\cite{Eff00}) we use the Lund string model FRITIOF~\cite{And93} in our 
model to describe high energy photoproduction on the nucleon. The particle 
production
in FRITIOF can be decomposed into two parts. First there is a 
momentum transfer taking place between the two incoming hadrons leaving two 
excited strings with the quantum numbers of the initial hadrons. After that
the two strings fragment into the observed particles. As a formation 
time we use 0.8~fm/c in the rest frame of each hadron; during this time 
the hadrons do not interact with the rest of the system. Since 
FRITIOF does not accept photons as incoming particles we use vector meson 
dominance (VMD)~\cite{Bau78} 
\begin{equation}
\label{eq:vmd}
  |\gamma\rangle=\left(1-\sum_{V=\rho,\omega,\phi}\frac{e^2}{2g_V^2}\right)|
\gamma_0\rangle+\sum_{V=\rho,\omega,\phi}\frac{e}{g_V}|V\rangle
\end{equation}
and pass the photon as a massless $\rho^0$, $\omega$ or $\phi$ with a 
probability corresponding to the strength of the vector meson coupling to the 
photon times its nucleonic cross section.\\
\begin{figure}[bt] 
  \begin{center}
    \vspace{-0.5cm}
    \includegraphics[width=10cm]{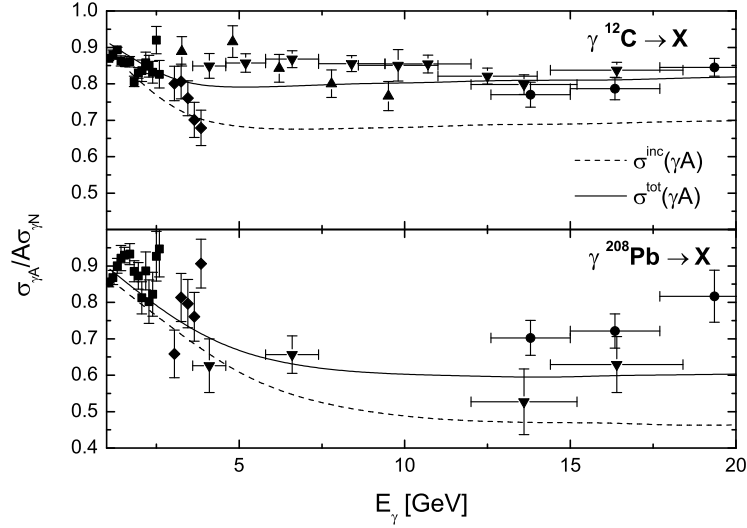}
  \end{center}
  \vspace{-0.5cm}
  \caption{The nuclear photoabsorption cross section $\sigma_{\gamma A}$ 
divided by $A\sigma_{\gamma N}$ plotted versus the photon energy $E_\gamma$. 
The solid line represents the result of Reference~\cite{Fal00} and the dashed 
line shows the incoherent part calculated using Eq.~(\ref{eq:intinc}). More 
than 90\% of the difference is due to coherent $\rho^0$ photoproduction.}
  \label{fig:abscs}
\end{figure}
In Glauber theory shadowing of incoherent meson photoproduction arises from
the interference between the two amplitudes depicted in 
Fig.~\ref{fig:Glauber}.
The left amplitude corresponds to the process where the photon directly 
produces the meson $X$ at nucleon $j$. The second amplitude of order 
$\alpha_{em}$ shows
the process where the photon first produces a vector meson $V$ on nucleon $i$ 
without excitation of the nucleus. This vector meson then propagates at fixed 
impact parameter $\vec b$ (eikonal approximation) to nucleon $j$ to produce 
the final state meson $X$ and leaving the nucleus in the same excited state as 
in the left amplitude. The FSI of the outgoing meson $X$ are usually treated
via a purely absorptive optical potential and lead to an exponential damping of
the nuclear production cross section 
$\sim\exp [-\sigma_{X}\int_{z_j}^\infty dz n(\vec b,z)]$. If one knew the 
amplitudes $T(\gamma N\rightarrow XN)$ and
$T(V N\rightarrow XN)$ one would just have to replace the purely absorptive
FSI by our transport model. However, these amplitudes are in general unknown.
To account for shadowing within our model we, therefore,
\begin{figure}[bt] 
  \begin{center}
    \vspace{-4cm}
    \includegraphics[width=14cm]{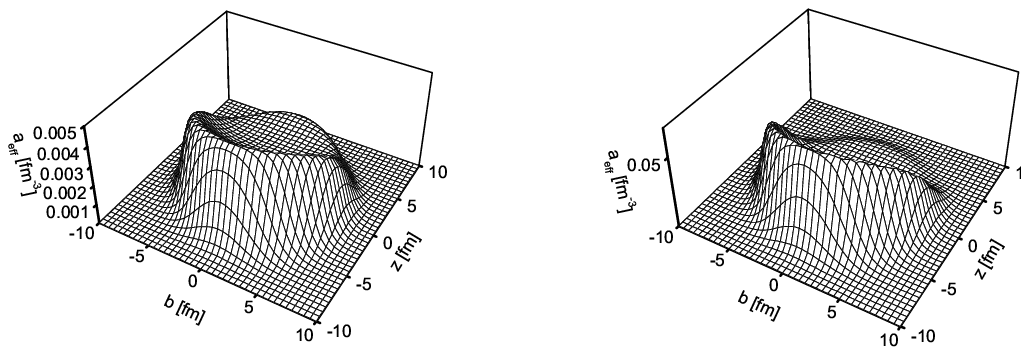}
  \end{center}
  \vspace{-2.5cm}
  \caption{The number density of nucleons that react with the $\phi$
component ({\it left side}) and $\rho^0$ component ({\it right side}) of a
20~GeV photon for $^{208}$Pb calculated using Eq.~(\ref{eq:aeffdens})
. In both 
cases the nucleons on the front side of the nucleus shadow the downstream 
nucleons. This effect is stronger for the $\rho^0$ component because of its 
larger nucleonic cross section.}
  \label{fig:aeff}
\end{figure}
start from (\ref{eq:vmd}) and use Glauber theory~\cite{Gla59Yen71} to calculate
how the single $V$ components of the photon change due to multiple scattering 
on the way to nucleon $j$ where the state $X$ is produced:
\begin{equation}
  \label{eq:gamma}
  |\gamma(\vec r_j)\rangle=\left(1-\sum_{V=\rho,\omega,\phi}\frac{e^2}{2g_V^2}\right)|\gamma_0\rangle+\sum_{V=\rho,\omega,\phi}\frac{e}{g_V}\left(1-\overline{\Gamma_V}(\vec r_j)\right)|V\rangle.
\end{equation}
Here $\Gamma_V\left(\vec r\right)$ denotes the (photon energy dependent) 
nuclear profile function.
The cross section for the photon to react with nucleon at position $\vec r$ 
inside the nucleus can be deduced via (\ref{eq:gamma}) from the optical theorem
\begin{equation}
\label{eq:inccs}
  \sigma_{\gamma N}(\vec r)=\left(1-\sum_{V=\rho,\omega,\phi}\frac{e^2}{2g_V^2}\right)^2\sigma_{\gamma_0 N}+\sum_{V=\rho,\omega,\phi}\left(\frac{e}{g_V}\right)^2\left|1-\overline{\Gamma_V}(\vec r)\right|^2\sigma_V.
\end{equation} 
Like for the photon in vacuum each term gives the relative
weight for the corresponding photon component to be passed to FRITIOF. 
When integrated over the whole 
nucleus one gets from Eq.~(\ref{eq:inccs}) the total incoherent photonuclear 
cross section
\begin{equation}
  \label{eq:intinc}
  \sigma_{\gamma A}^{inc}=\int d^3r_jn(\vec r_j)\sigma_{\gamma N}(\vec r_j)
\end{equation}
which is shown in Fig.~\ref{fig:abscs} together with the total nuclear 
photoabsorption cross section as calculated in~\cite{Fal00}. More than 90\% of 
the difference between those two cross sections stems from coherent $\rho^0$ 
photoproduction. In Fig.~\ref{fig:aeff} we show how strongly the $\rho^0$ and
the $\phi$ component of a real 20~GeV photon are shadowed in Pb. We plot
\begin{equation}
  \label{eq:aeffdens}
  a_{eff}^V(\vec r_j)=n(\vec r_j)\frac{1}{\sigma_{\gamma N}}\left(\frac{e}{g_V}\right)^2\left|1-\overline{\Gamma_V}(\vec r_j)\right|^2\sigma_V
\end{equation}
as a function of $\vec r_j$. One clearly sees that due to its smaller 
nucleonic cross section the $\phi$ component is less shadowed than the
$\rho^0$ component at the backside
of the nucleus. This means that strangeness production (e.g. $K$ production), 
where the primary reaction is preferably triggered by the $\phi$ component of 
the photon, is less shadowed than for instance inclusive $\pi$ production. This
dependence of the strength of shadowing on the reaction type is new 
compared to the shadowing in~\cite{Eff00} and can also be seen directly from 
the second amplitude in Fig.~\ref{fig:Glauber} because of the occurrence of 
the scattering process $VN\rightarrow XN$ at nucleon $j$.\\

As already mentioned above the purely absorptive FSI of the Glauber model
are very different from the coupled channel description of a transport model.
The transport model we use is based on the BUU equation that describes the time
evolution of the phase space density $f(\vec r,\vec p,t)$ of particles
that can interact via binary reactions. In our case these particles are the 
nucleons of the target nucleus as well as the baryonic resonances and mesons 
($\pi$, $\eta$, $\rho$, $K$, ...) that can either be produced in the primary 
$\gamma N$ reaction or during the FSI. For particles of type $i$ the BUU 
equation looks as follows:
\begin{equation}
  \left(\frac{\partial}{\partial t}+\frac{\partial H}{\partial\vec r}\frac{\partial}{\partial \vec r}-\frac{\partial H}{\partial \vec r}\frac{\partial}{\partial \vec p}\right)f_i(\vec r,\vec p,t)=I_{coll}[f_1,...f_i,...,f_M].
\end{equation}
In the case of baryons the Hamilton function $H$ includes a mean field 
potential which in our model depends on the particle position and momentum. 
The collision integral on the right hand side accounts for the creation and
annihilation of particles of type $i$ in a collision as well as elastic 
scattering from one position in phase space into another. For fermions Pauli 
blocking is taken into account in $I_{coll}$ via Pauli factors. For each 
particle type $i$ such a BUU equation exists; all are coupled via the mean 
field and the collision integral. This leads to a system of coupled 
differential-integral equations which we solve via a test particle ansatz for 
the phase space density
\begin{equation}
  f_i(\vec r,\vec p,t)=\frac{1}{N}\sum_j\delta(\vec r-\vec r_j)\delta(\vec p-\vec p_j).
\end{equation}
\begin{figure}[bt] 
  \begin{center}
    \includegraphics[width=10cm]{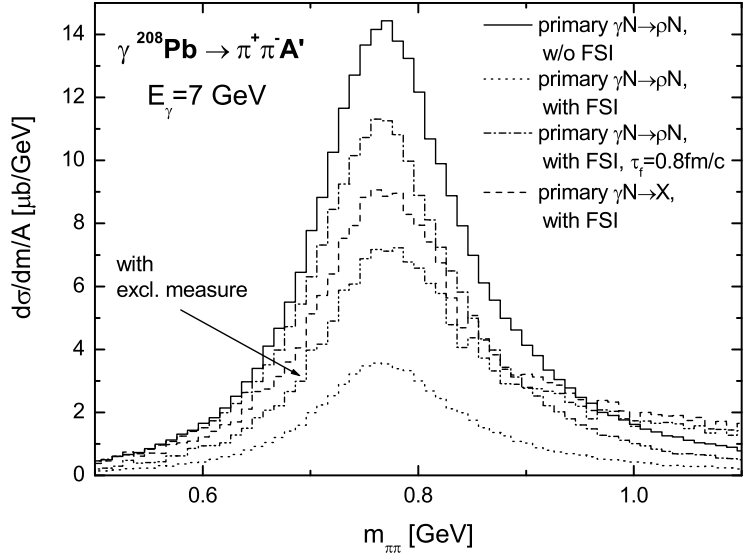}
  \end{center}
  \vspace{-0.5cm}
  \caption{Mass differential cross section for exclusive $\rho^0$ production
on $^{208}$Pb at $E_\gamma=7$~GeV. The meaning of the different curves is 
explained in detail in the text. All the curves, except the one with the
explicitly given formation time $\tau_f=0.8$~fm/c have been calculated with
$\tau_f=0$.}
  \label{fig:exclrho}
\end{figure}
For a system of non interacting particles ($I_{coll}=0$) this leads
directly to the classical equations of motion for the test particles:
\begin{equation}
  \frac{d\vec r_j}{dt}=\frac{\partial H}{\partial \vec p_j}\qquad\frac{d \vec p_j}{dt}=-\frac{\partial H}{\partial \vec r_j}.
\end{equation}
Since the collision integral also accounts for particle creation in a collision
the observed outgoing particle $X$ cannot only be produced in the primary 
reaction but can also be created by side feeding in which a particle $Y$ is 
created first which propagates and then, by FSI, produces $X$. In addition the 
state $X$ might get absorbed on its way out of the nucleus but be fed in again 
in a later interaction. Both cases can a priori not be ignored but are usually 
neglected in Glauber models.

\section{Results}
Exclusive vector meson photo- and electroproduction on nuclei is an ideal tool 
to study the effects of the coherence length, formation time and color 
transparency. It
has been investigated at HERMES~\cite{Ack00} at photon energies between 10~GeV 
and 20~GeV and $Q^2\lesssim$~5~GeV$^2$. 
\\

The calculations for meson production
on nuclei are usually done within simple Glauber theory~\cite{Glauber-models}.
As already mentioned above the FSI in Glauber theory are usually purely 
absorptive. This means that for the reaction $\gamma A\rightarrow\rho^0 A^*$
the primary reaction has to be $\gamma N\rightarrow\rho^0 N$. If one treats the
FSI via an absorptive optical potential one gets an exponential damping
$\sim\exp [-\sigma_{\rho N}\int_{z_j}^\infty dz n(\vec b,z)]$ of the nuclear
production cross section. Up to now we can, for technical reasons, perform
calculations only for real photons up to an energy of about 7~GeV. In 
Fig.~\ref{fig:exclrho} we show the results for the mass differential cross
section of incoherent $\rho^0$ photoproduction on $^{208}$Pb for 
$E_\gamma=7$~GeV. The solid line represents a calculation where the primary 
reaction is $\gamma N\rightarrow\rho^0N$. It already includes the effects
of shadowing, Fermi motion, Pauli blocking and the nucleon potential, but no 
FSI. The dotted line shows the effect of the FSI without a formation time of
the $\rho^0$ in $\gamma N\rightarrow\rho^0N$. The simple Glauber model yields
exactly the same result, which means that FSI processes like 
($\rho^0N\rightarrow\pi N$, $\pi N\rightarrow\rho^0N)$ where the primary 
$\rho^0$ gets absorbed first and is fed into the outgoing channel by a later 
FSI are negligible. If one assumes a formation time of $\tau_f=0.8$~fm/c for 
the $\rho^0$ one gets the result indicated by the dash-dotted line. Due to the
finite formation time there is less absorption and the nuclear production cross
section increases. If the spectrum looked like this one would in Glauber 
theory be lead to the conclusion of a finite $\rho^0$ formation time.\\ 

However,
one will get a similar result with $\tau_f=0$ if one allows for other primary 
reaction besides $\gamma N\rightarrow\rho^0N$ and uses a coupled channel model.
This can be seen by looking at the dashed line in Fig.~\ref{fig:exclrho}. In
this case one finds that the main part of the additional $\rho^0$ stem from
inclusive $\rho^0$ production in the primary event, e.g. 
$\gamma N\rightarrow\rho^0\pi N$ where the $\pi$ gets absorbed during the FSI.
One could now apply an exclusivity measure like in the HERMES experiment
\begin{equation}
  \Delta E=\frac{P_Y^2-M_N^2}{2M_N},
\end{equation}
where $P_Y$ denotes the 4-momentum of the undetected final state and $M_N$ is
the nucleon mass, with -2~GeV$<\Delta E<$0.6~GeV. 
\begin{figure}[bt] 
  \begin{center}
    \includegraphics[width=10cm]{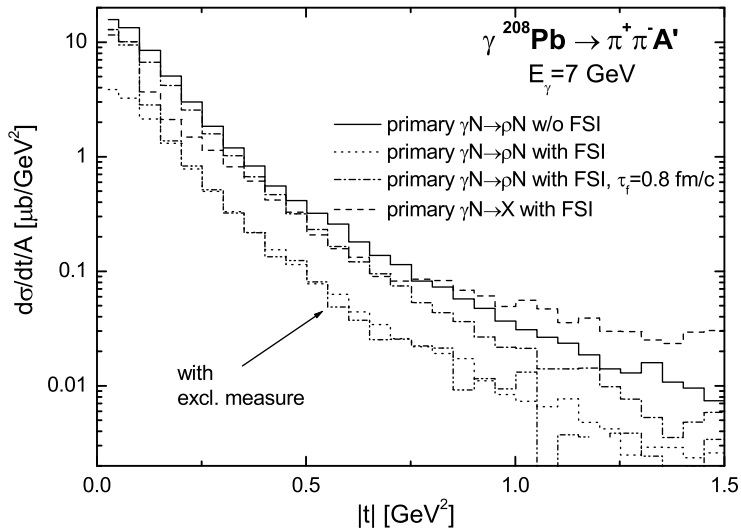}
  \end{center}
  \vspace{-0.5cm}
  \caption{Calculated $\frac{d\sigma}{dt}$ for exclusive $\rho^0$ production
on $^{208}$Pb at $E_\gamma=7$~GeV. The meaning of the different curves is the 
same as in Fig.~\ref{fig:exclrho}.}
  \label{fig:exclrho2}
\end{figure}
This leads to a decrease of the cross section (dash-dot-dotted line) because 
some of the inclusive primary events are excluded. If the exclusivity measure 
was good enough to single out only the exclusive primary events,
the curve would coincide with the dotted line and Glauber theory would be 
applicable. Since this is not the case, one still extracts an incorrect
formation time when using simple Glauber theory. 
One therefore needs an additional constraint 
to make Glauber theory applicable. This becomes clear by examining the
differential cross section $\frac{d\sigma}{dt}$ in Fig.~\ref{fig:exclrho2}. The
meaning of the lines are as before. One can see that for $|t|>0.1$~GeV$^2$ the
full calculation with exclusivity measure (dash-dot-dotted line) gives the
same result as the one with the primary reaction $\gamma N\rightarrow\rho^0N$
and FSI (dotted line). This means that only in this kinematic region Glauber
theory can be used. In the HERMES experiment one makes a lower $|t|$ cut to
get rid of the $\rho^0$ stemming from coherent $\rho^0$ photoproduction. In
the case of lead and $E_\gamma=7$~GeV the coherent part can be neglected above 
$|t|=0.05$~GeV$^2$. If one wants to apply Glauber theory one will need to 
increase this threshold to approximately $|t|=0.1$~GeV$^2$ to suppress 
contributions from inclusive $\rho^0$.\\
\begin{figure}[bt]
  \begin{center}
    \vspace{-2cm}
    \includegraphics[width=14cm]{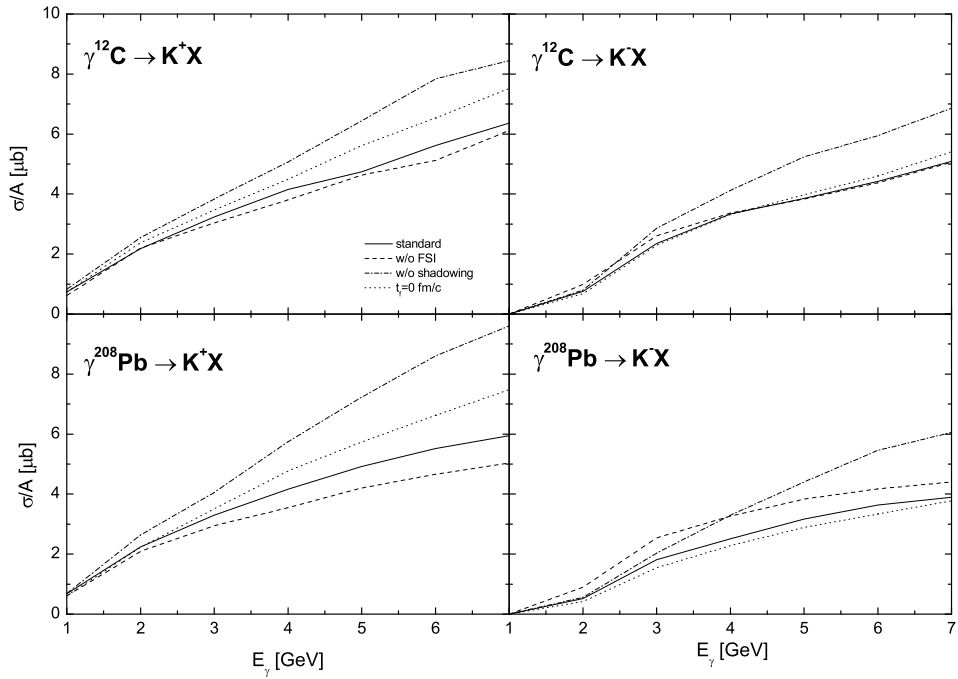}
  \end{center}
  \vspace{-1cm}
  \caption{Photoproduction cross section for $K^+$ (left) and $K^-$ (left) for 
$^{12}$C and $^{208}$Pb plotted as a function of the photon energy. The solid
line represents the full calculation. The dash-dotted line shows the result 
without shadowing of the incoming photon, the dashed line the result without
FSI and the dotted line the calculation without formation time.}
  \label{fig:kaons}
\end{figure}

One sees that Glauber theory can be trusted only under certain kinematic 
constraints. To 
describe less exclusive reactions one needs a more realistic description of 
the FSI. In Fig.~\ref{fig:kaons} we show the cross section for the reactions
$\gamma A\rightarrow K^+X$ and $\gamma A\rightarrow K^-X$ in the photon energy
range 1-7~GeV for $^{12}$C and $^{208}$Pb which has already been investigated 
in~\cite{Eff00}. The solid curve in Fig.~\ref{fig:kaons} represents the 
results of the full calculation (including shadowing, FSI, $\tau_f=0.8$~fm/c, 
etc.). By comparison with the calculation without shadowing (dash-dotted line)
one sees how important shadowing becomes at high energies. At 7~GeV it reduces 
the nuclear  production cross section to about 75\% for Carbon and 65\% for 
Lead. The importance of a full coupled channel treatment of the FSI becomes 
clear 
when looking at inclusive $K^+$ production. Since the $\overline{s}$ quark
cannot be absorbed in medium the FSI can just increase the $K^+$ production
via processes like $\pi N\rightarrow K^+Y$ ($Y=\Sigma,\Lambda$) for example.
This one finds by comparison with the calculation without FSI (dashed line). 
As a consequence of this, a shorter formation time will lead to an 
increase of the $K^+$ production cross section as can be seen from the dotted
line. The reason is that with decreasing formation time the primarily produced
pions have a greater chance to produce $K^+$ in the FSI. As already mentioned
in the introduction for $\tau_f=0.8$~fm/c the formation length of fast pions is
larger than the nuclear dimension, so that they will leave the nucleus without
further scattering. An enhancement of the $K^+$ production cross
section due to FSI can of course not be explained by purely absorptive FSI as 
in simple Glauber theory. The $K^-$ can also be absorbed via processes like 
$K^-N\rightarrow\pi Y$. This effect compensates the production due to FSI in 
$^{12}$C and dominates in $^{208}$Pb as can bee seen in Fig.~\ref{fig:kaons}.

\section{Summary \& Outlook}
We have shown that high energy photoproduction off nuclei offers a great 
possibility to study the physics of hadron formation. However, one needs a
reliable model of the FSI to extract the formation time from the production
cross sections. Whereas Glauber models allow for a straight forward
implementation of the nuclear shadowing effect they usually have the 
disadvantage of a purely absorptive treatment of the FSI. As we have shown the
latter might lead to a wrong estimate of the formation time if the 
kinematic cuts are
not chosen properly. A more realistic treatment of the FSI is possible within 
a coupled channel transport model. However in such a model it is not totally
clear how to account for the shadowing effect. We have presented a method to 
account for these
coherence length effects which can easily be extended to higher energies and 
virtual photons. This will be done in future work so that we can study the
effects of formation time and color transparency in particle
production in the HERMES regime~\cite{Ack00,HERMES01}. We will also investigate
charm production since we expect the same FSI effect as observed for the $K^+$
also for the $\overline{D}$. In addition, we plan to implement a more realistic
time dependence of the cross section for the produced particles during their 
evolution which might again modify the extracted formation time.

\section{Acknowledgments}
This work was supported by DFG.


\end{document}